\documentclass{cidr-2020}

\usepackage[utf8]{inputenc}
\usepackage[english]{babel}

\usepackage{blindtext}
\usepackage{amssymb}

\usepackage{graphicx}
\usepackage{balance}  
\usepackage{xcolor}

\usepackage{balance}
\usepackage{graphicx, comment}
\usepackage{cite}
\usepackage{url}

\usepackage{breakurl}
\usepackage{balance}  
\usepackage{booktabs} 
\usepackage{tabularx}
\usepackage{algorithm, algorithmicx, algpseudocode}
\usepackage{pifont}
\usepackage{xspace}
\usepackage{caption}
\usepackage{subcaption}

\newcommand{\framework}{TXSC\xspace}

\makeatletter
\def\@copyrightspace{\relax}
\makeatother

\begin{document}

\title{Transactional Smart Contracts in Blockchain Systems}

\numberofauthors{3}
\author{
%
%
\alignauthor
Victor Zakhary\\
       \affaddr{UC Santa Barbara}\\
       \affaddr{Santa Barbara, California} \\
       \affaddr{USA, 93106}\\
       \email{victorzakhary@ucsb.edu}
\alignauthor
Divyakant Agrawal\\
\affaddr{UC Santa Barbara}\\
       \affaddr{Santa Barbara, California} \\
       \affaddr{USA, 93106}\\       \email{divyagrawal@ucsb.edu}
\alignauthor
Amr El Abbadi\\
	   \affaddr{UC Santa Barbara}\\
       \affaddr{Santa Barbara, California} \\
       \affaddr{USA, 93106}\\       
       \email{elabbadi@ucsb.edu }
}

\maketitle

\begin{abstract}
This paper presents \framework, a framework that provides smart contract developers
with transaction primitives. These primitives allow developers to write smart contracts without the need to reason about the anomalies that can arise due to concurrent smart contract function executions.
\end{abstract}

 \section{Introduction}

Executing concurrent operations has been a long-term challenge in the 
design of large software systems. Without careful usage of 
synchronization primitives~\cite{dijkstra1968cooperating}, the concurrent
execution of multiple procedures that access shared variables can easily
result in anomalous executions. Instead of using synchronization 
primitives, that a programmer must carefully program, database systems
introduced the elegant declarative notion of {\em 
transactions}~\cite{gray1981transaction}. Programs 
that may be executed concurrently are each executed as a transaction, and the 
database management system ensures that transaction execution is isolated from each
other and that the concurrent and interleaved execution of multiple transactions is
serializable, i.e., equivalent to a serial execution~\cite{bernstein1987concurrency}. 

Recent interest in blockchains has resulted in its rapid usage in diverse 
applications, and its evolution to support complex concurrent executions.  The 
original blockchain, as proposed in Bitcoin\cite{nakamoto2008bitcoin}, involved simple 
\emph{transactions}, that transfer some bitcoins from one end-user (typically 
Alice) to another end-user (typically Bob).  The original bitcoin blockchain can be
easily modelled as an abstract data type representing a linked list of blocks of 
transactions.  The accessed data is the cryptocurrency, bitcoins, and transactions 
transfer part of the remaining, unused assets of Alice to Bob, while keeping the 
rest with Alice (hence the term Unspent Transaction Output, UTXO to refer to the 
assets belonging to a client in Bitcoin).  A {\em miner} adds a transaction to a 
block if the assets consumed in the transaction are not double spent in the same 
block and if the miner can validate that the end-user does actually have these 
assets, i.e., the UTXO actually belongs to the end-user issuing the transaction.  
Finally, a miner adds a block to the blockchain if it solves the Proof of Work (PoW)
puzzle~\cite{nakamoto2008bitcoin}.    

Ethereum~\cite{wood2014ethereum} reintroduced the notion of {\em smart contracts}~\cite{szabo1997formalizing} to blockchains.
Smart contracts extend the simple abstract data type notion of blockchain transactions to 
include complex data type classes with end-user defined variables and functions.  
When an end-user deploys a smart contract in a blockchain, this deployment results 
in instantiating an object instance of the smart contract class in the 
blockchain~\cite{herlihy2019blockchains,dickerson2017adding}. 
The object state is initially stored in the block where the object is instantiated.
End-users can issue a smart contract function call by sending function call 
requests to the miners of a blockchain. These function calls are transactions that 
are sent to the \emph{address} of the smart contract object.  Miners execute these 
transactions and record object state changes in their currently mined block. 
Therefore, the state of a smart contract object could span one or more blocks of a 
blockchain.

Smart contracts now have their own variables and multiple functions that may be 
executed by different end-users results in transactions which might be incorporated
in different blocks by different miners. This clearly results in complex 
concurrency challenges which need to be handled by smart contract developers. 
Distributed database literature~\cite{corbett2013spanner,shute2012f1} has shown that 
putting the burden of implementing transaction logic in the application layer is 
problematic. This is no simple task and serious smart contract 
concurrency bugs have been highlighted in the
blockchain literature~\cite{kolluri2018exploiting, luu2016making, sergey2017concurrent, dickerson2017adding}.  In fact, from a financial point-of-view, 
two such famous anomalies in the context of blockchains,   TheDAO~\cite{TheDAOOrganization, TheDAOBug} and the BlockKing~\cite{BlockKingContract} have resulted in a loss of tens of
millions of investors' dollars~\cite{luu2016making}.

In this paper, we advocate leveraging the traditional transactional approach to 
address the concurrency violations in the context of smart contract executions in 
large scale blockchain systems. In particular, we propose Transactional Smart 
Contracts (\framework) as a framework that allows developers to write smart 
contracts with correct transaction isolation semantics. Unlike previous 
works~\cite{kolluri2018exploiting, luu2016making, sergey2017concurrent} that 
propose smart contract analysis tools to detect concurrency bugs in smart 
contracts, \framework aims to free smart contract developers from the burden of 
implementing correct concurrency control semantics for each smart contract. 
Instead, developers can focus on the smart contract application semantics and leave
the concurrency semantics to \framework.

Concurrency control problems arise in two general contexts during smart contract 
function execution depending on whether the application semantic 
functionality is implemented by a single or multiple functions.  In a \textbf{single 
function}, each function in a smart contract is executed correctly (and in 
isolation) as a miner validates its execution.  However, the state of the data in 
the blockchain is visible and can be read all the time by any end-user.  An 
end-user might take action based on a value read, but due to the concurrent 
execution of smart contract functions, such a read value might be stale when the 
function is executed.  \framework needs to ensure that the attribute values observed 
by an end-user, where these attributes are in the read set of a function, are still valid when the function is executed. Alternatively, the 
semantic functionality might be executed by \textbf{multiple functions} in the same or even
different smart contracts on potentially different blockchains.  These functions 
might invoke each other in an asynchronous manner.  In particular, a function, 
before termination may call another function to perform a specific task, which in 
turn calls a third function, and so on.  This arises due to smart contracts in a 
single blockchain like the puzzle example in~\cite{luu2016making} or 
across multiple chains~\cite{BlockKingContract, TheDAOBug} that requires atomic execution across blockchains~\cite{atomicNolan,herlihy2018atomic,zakhary2019atomic}.  
In this case, different invocations of the function might be interleaved
resulting in incorrect executions due to the lack of isolation.

In this paper, we propose the Transactional Smart Contracts paradigm to solve these concurrency problems.  In particular, 
\begin{enumerate}
    \item This paper models smart contract concurrency anomalies as transaction isolation problems. Examples illustrate how different smart contract concurrency anomalies can be mapped to the problem of transaction isolation of either single domain or distributed cross-domain transactions.
    \item \framework is the first framework to provide smart contract developers with transactional primitives \textit{start transaction} and \textit{end transaction}.  \framework takes a smart contract that contains these primitives as an input and translates it to a transactionally correct smart contract using the smart contract native language.
\end{enumerate}
The rest of the paper is organized as follows. We start with two examples to 
illustrate the types of concurrency anomalies that can arise in the context of 
smart contracts in Section~\ref{sec:examples}.
Data and transaction models are presented
in Section~\ref{sec:models}. Section~\ref{sec:serializable} explains our solution 
and  presents \framework and the paper is concluded in Section~\ref{sec:conclusion}.

 \section{Concurrency Anomalies in Smart Contracts} \label{sec:examples}
Most of the smart contract anomalies identified in prior work~\cite{kolluri2018exploiting, luu2016making, 
sergey2017concurrent, dickerson2017adding} are rooted to faulty transaction isolation semantics
implemented by the smart contract developers. These anomalies can be classified into \textit{two} categories: 1) faulty
transaction isolation semantics among transactions that span a single administrative domain (or one blockchain) and
2) faulty transaction isolation semantics among distributed transactions that span several administrative
domains (more than one blockchain or one blockchain and services outside the domain of this blockchain). We explain
the two categories using the following two examples from~\cite{luu2016making} and~\cite{sergey2017concurrent}.
For consistency with the original blockchain terminology, in this section, we refer to a 
function call request as a transaction (later we will change this).

\textbf{The puzzle example.} This example illustrates the first category of smart contract concurrency anomalies.
In this example, an end-user, \textit{the challenger}, deploys a smart contract that pays another end-user, 
\textit{the solver}, a reward if the solver's submitted puzzle solution
is correct. Algorithm~\ref{algo:puzzle-smart-contract} shows the puzzle smart contract pseudocode. As shown, the smart contract
has three functions: a {\tt Constructor} (Line~\ref{line:puzzle-constructor}), {\tt UpdateReward} 
(Line~\ref{line:puzzle-update-reward}), and {\tt SubmitSolution} (Line~\ref{line:puzzle-submit-solution}) functions. The 
{\tt Constructor} is executed by the contract owner, the challenger, to initialize the smart contract object. {\tt UpdateReward} can 
be executed only by the challenger to update the reward value of the puzzle. Furthermore, {\tt UpdateReward} can only be executed if the puzzle
has not been solved yet (Line~\ref{line:puzzle-issolved}) and {\tt UpdateReward} sends the old reward value to the challenger and
updates the reward value with the new value sent by the challenger 
(Line~\ref{line:puzzle-set-reward}). {\tt SubmitSolution} 
(Line~\ref{line:puzzle-submit-solution}) allows any solver to submit a solution to the 
puzzle only if the puzzle has not been solved yet. If the submitted solution is
correct (Line~\ref{line:puzzle-correct-solution}), the reward goes to the solver, the
puzzle's solution is updated, and the puzzle is marked as solved.

Now, assume Alice is a challenger who posts a puzzle that follows the smart contract 
description in Algorithm~\ref{algo:puzzle-smart-contract} in the Ethereum network and 
she sets the reward value $r$ to $r = 2$ ethers, the currency of the Ethereum network. 
Bob, a solver, reads the reward value $r = 2$ ethers, solves the puzzles, and submits 
the solution to the smart contract through a transaction $TX_1$. Bob assumes to receive 
a puzzle reward of 2 ethers if his solution is correct. Concurrently, Alice
might, benignly or maliciously, schedule a transaction $TX_2$ that updates the reward of 
the puzzle to a smaller value than the current reward e.g., $r = 0$.  If $TX_2$ is 
executed first, $r$ would be updated to its new value $0$. While updating the 
reward value should result in aborting $TX_1$ as the value of $r$ read by $TX_1$ is
stale, the smart contract code in Algorithm~\ref{algo:puzzle-smart-contract} would
allow $TX_1$ to execute. This results in Alice receiving a solution to her puzzle 
while Bob gets a reward of 0 ethers. 
As both $TX_1$ and $TX_2$ access an object that 
spans only one blockchain, the Ethereum network, this concurrency anomaly falls into
the first category of the two aforementioned categories.

\begin{algorithm}[t]
\caption{Puzzle smart contract example in~\cite{luu2016making}}
 \label{algo:puzzle-smart-contract}

{\sf class} Puzzle \{
    
\begin{algorithmic}[1]
    \State {\sf address public} owner  \Comment{{\tt contract owner}}
    \label{line:puzzle-contract-owner}
    \State {\sf bool public} solved \Comment{{\tt true if the puzzle is solved}}
    \State {\sf uint public} reward \Comment{{\tt puzzle solving reward}}
    \State {\sf bytes32 public} diff     \Comment{{\tt puzzle difficulty}}
    \State {\sf byte32 public} solution    \Comment{{\tt puzzle solution if found}}
    \label{line:car-owner}
    
    \Procedure{Constructor}{} \label{line:puzzle-constructor}
        \State {\sf this}.owner = msg.sender
        \State {\sf this}.reward = msg.value
        \State {\sf this}.solved = false
        \State {\sf this}.diff = bytes32(msg.data) \Comment{{\tt set difficulty }}
    \EndProcedure
    
    \Procedure{UpdateReward}{} \label{line:puzzle-update-reward}
        \State {\sf requires}($msg.sender == this.owner$) \label{line:puzzle-owner-requirement}
        \If {$!$ solved} \label{line:puzzle-issolved}
        \State transfer reward to owner
        \State reward = msg.value \label{line:puzzle-set-reward}
        \EndIf
    \EndProcedure
    
    \Procedure{SubmitSolution}{} \label{line:puzzle-submit-solution}
        \If {$!$ solved}
        \If {sha256(msg.data) < diff} \label{line:puzzle-correct-solution}
        \State transfer reward to msg.sender
        \State solution = msg.data  
        \State solved = true
        \EndIf
        \EndIf
    \EndProcedure
    
\end{algorithmic}
\}
\end{algorithm}

 \textbf{The BlockKing~\cite{sergey2017concurrent,BlockKingContract} example.} This example
 demonstrates the second category of smart contract concurrency 
 anomalies where end-user 
 distributed transactions span several administrative domains (objects of one or
 more blockchains in addition to asynchronous calls to external services). Algorithm~\ref{algo:blockking-smart-contract} shows
 code snippets from the original 366 lines of code of the BlockKing smart 
 contract~\cite{BlockKingContract} where concurrency anomalies occur. The BlockKing smart 
 contract works as follows. At any 
 moment in time, there exists one block king, initially, the contract owner. Users send 
 money to the contract via the {\tt Enter} function (Line~\ref{line:BlockKing-enter}) as bids 
 to become the next block king. The {\tt Enter} function stores the address of the
 caller, the current block number, and the caller's bid value in the attributes 
 {\tt warrior}, {\tt warriorBlock}, and {\tt warriorGold} respectively. Then, the {\tt Enter} 
 function calls an external random number generator to generate a random number 
 between 1-9 and if the returned number equals to the first digit of the block number
 stored in the {\tt warriorBlock} attribute, the caller of the {\tt Enter} function 
 becomes the new block king. A block king gets a percentage of the bid money of every
 call to the {\tt Enter} function and the contract owner gets the remaining 
 percentage of this bid money. Notice that the random number generator triggers an 
 asynchronous callback function (Line~\ref{line:BlockKing_callback}) where the 
 returned random number is checked against the block number in the {\tt warriorBlock} 
 attribute. If the returned random number matches
 the first digit of the block number in the {\tt warriorBlock}, the current warrior becomes
 the new block king.
 
 \begin{algorithm}[t]
\caption{Snippets from the BlockKing contract~\cite{BlockKingContract}}
 \label{algo:blockking-smart-contract}

{\sf class} BlockKing \{
    
\begin{algorithmic}[1]
    \State {\sf address public} king, warrior
    \State {\sf uint public} kingBlock, warriorBlock
    \State {\sf uint public} warriorGold, randomNumber
    
    \Procedure{enter}{} \label{line:BlockKing-enter}
        \State ... \Comment{{\tt check if minimum bet is sent}}
        \State warrior = msg.sender, warriorGold = msg.value
        \State warriorBlock = block.number
        \State byte32 myid = oraclize\_query(0, "WolframAlpha", "random number between 1 and 9")
    \EndProcedure
    
    \Procedure{\_callback}{byte32 myid, string result} \label{line:BlockKing_callback}
        \State requires(msg.sender ==  oraclize\_cbAddress())
        \State randomNumber = uint(bytes(result)[0]) - 48;
        \If {singleDigitBlock == randomNumber}
        \State ... \Comment{{\tt update reward}}
        \State king = warrior, kingBlock = warriorBlock
        \EndIf
        
    \EndProcedure
    
\end{algorithmic}
\}
\end{algorithm}
If calls to the {\tt Enter} function are blocking; meaning that at
most one call to the {\tt Enter} function is allowed until its callback is 
completed, the smart contract in 
Algorithm~\ref{algo:blockking-smart-contract} would not have any 
concurrency anomalies.  However, the smart contract in 
Algorithm~\ref{algo:blockking-smart-contract} is non-blocking. This non-blocking
behavior allows many concurrent calls to the {\tt Enter} function to take place. 
If multiple transactions are concurrently sent to the {\tt Enter} function, each 
transaction would replace the values of the {\tt warrior}, the {\tt warriorBlock}, and the 
{\tt warriorGold} attributes of all the previous incomplete transactions. This leads to an 
advantage to the latest caller who sends a transactions to the {\tt Enter} function 
before all previous callbacks occur. Every trigger to the callback function 
gives the latest caller a chance to become the new block king while previous callers 
have no chance to become the new block king. We illustrate this transaction isolation 
anomaly using the following example. Assume Alice, Bob, and Carol 
concurrently want to become the next block king. They send three transactions (corresponding
to three {\tt Enter} function calls)
$TX_1$, $TX_2$, and $TX_3$ accompanied by their bids to the \textit{enter} function
respectively. $TX_1$ updates the warrior attributes to Alice's attributes sent along 
with $TX_1$ then, calls the external random number generator. Before $TX_1$'s callback 
is triggered, $TX_2$ replaces the warrior attributes with Bob's attributes sent with 
$TX_2$ and similarly, $TX_3$ replaces the warrior attributes with Carol's attributes.
When the callbacks of $TX_1$, $TX_2$, and $TX_3$ are triggered, which possibly could 
take place in another block in the BlockKing blockchain, the three callbacks use the
warrior attribute values of Carol to decide if she could be the next block king or not.
Carol gets 3 chances to become the block king while Alice and Bob have no chance. 

This concurrency violation occurs as transactions $TX_1$, $TX_2$, and $TX_3$ are not being
executed in isolation.

\textbf{Transactions in the first category can be 
atomically executed in one shot within one block of its smart contract blockchain. On
the other hand, distributed transactions could span multiple blocks in one or more 
blockchains and hence ensuring their atomicity while executing them in isolation is 
significantly more complicated than executing transactions in the first category in 
isolation.}

\section{Data and Transaction Models} \label{sec:models}

An open permissionless blockchain~\cite{maiyya2018database} comprises
an application layer and a storage layer. Clients in the application
layer have public identities represented by their public keys and 
private signatures generated using their private keys. Clients send
signed transactions to the storage layer in order to transfer assets
from one client to another.  The storage layer consists of mining or 
computing nodes, \textit{miners}, and each miner manages a copy of 
the blockchain. Transactions, in the storage layer, are grouped into 
blocks and each block is hash chained to the previous block; hence 
the name \textit{blockchain}. When a mining node receives a 
transaction, it verifies the transaction and adds it to its current 
block, \textit{only if the transaction is valid}. Mining nodes run a consensus algorithm or in a permissionless
blockchain Proof of Work (PoW) to reach consensus on the next 
block to be added to the blockchain.

Smart contracts are analogous to classes~\cite{herlihy2019blockchains,dickerson2017adding,zakhary2019towards} in Object Oriented 
Programming Languages (OOPL) and are used by clients to implement
complex data types. Clients deploy smart contracts to a
blockchain by sending a deployment message to miners of this blockchain. 
As a result, a miner instantiates an object of the smart
contract class and stores this object in the current block in the blockchain. 
Smart contract objects have attributes that capture their state. Once a smart
contract object is instantiated in a blockchain, the state of
this object, as part of the blockchain, is made public and can be 
\textbf{externally read by any client at any moment}. In addition, smart contract 
objects have functions that define the possible state transitions of these objects. Since an
object state is public, smart contract read-only functions are pointless. 
Therefore, it is
safe to assume that any smart contract function call has to update at least one
attribute of the smart contract object~\cite{wustace}. A smart 
contract object has an address in the blockchain. When a client wants to issue 
a smart contract function call, the client sends a function call request to the
miners of the blockchain where the smart contract is deployed. This function 
call request is directed to the address of the smart contract object. Miners 
use the smart contract address to locate the smart contract object (state and code).
This function call is accompanied by some implicit 
parameters like \textit{msg.sender}, the address of the client who sent the 
transaction, \textit{msg.val}, the value of the money sent along with the 
transaction, and \textit{msg.data}, any data that needs to be sent along with
the transaction. In addition, function calls could be accompanied by some
function explicit parameters.

We follow the Ethereum~\cite{wood2014ethereum} smart contract execution model.
Each function call is accompanied by some $gas$ value. The $gas$ value represents
the amount of money a client is willing to pay to incentivize miners to execute 
the function call. Miners charge some $gas$ for every executed line of 
code in the called function. A miner stores any intermediate results of a function
call in their local storage. If the function call completes before the function
call runs out of
$gas$, the intermediate results are finalized and included in the miner's current
block. However, if a function call runs out of $gas$ before the function call
is completed, intermediate results are deleted and the smart contract object 
state does not change. Either way, the miner includes a transaction that pays the
miner the 
amount of $gas$ spent during the execution of the function call in its current 
block. Smart contract function calls are atomic meaning that each function call
either terminates after it successfully updates the object state in the blockchain or 
rolls back to the object state before the call occurs. Concurrent function calls are 
sequentially executed one after the other without any 
interruption~\cite{sergey2017concurrent}.  In blockchain terminology, a 
function call request is usually referred to as a transaction. Yet, a function 
call might not ensure the ACID~\cite{bernstein1987concurrency} properties of
transactions in traditional databases. 

In traditional DBMS, a client transaction starts when a client calls the
\textit{start (begin) transaction} command. Afterwards, a transaction 
reads and updates some data values followed by an \textit{end (commit) 
transaction} command. The role of the DBMS is to ensure the ACID properties
of a client transaction from the moment the transaction begins till the moment
the transaction ends (whether the transaction commits or aborts).

In permissionless blockchains, miners have no way to learn the details of all
client activities before calling the smart contract functions, e.g., when the client activities start and what 
values were read before a function call request is sent to the miners. Even when each 
function call is executed
in isolation from  concurrent function calls, transaction isolation 
concurrency violation still occur as shown in 
Algorithms~\ref{algo:puzzle-smart-contract} 
and~\ref{algo:blockking-smart-contract} as a result of poor client 
transaction isolation, network asynchrony, and smart contract 
asynchronous callbacks. We consider a {\em client transaction span} to 
include all the read operations that took place before the client sends a
function call, the function execution caused by the function call, and 
any callbacks that are triggered as a result of this function call.
The goal of this paper is to ensure the ACID properties of client 
transactions from the time a client starts a transaction till the end of 
the function call that terminates this transaction.

\section{Transactional Smart Contracts} \label{sec:serializable}

\begin{algorithm}[ht!]
\caption{A smart contract example that uses  \framework}
 \label{algo:example-smart-contract}

{\sf class} SmartContract
    
\begin{algorithmic}[1]
    \Procedure{f1}{} \label{line:example-f1}
        \State start transaction
        \State f1's logic
        \State end transaction
    \EndProcedure
    
    \Procedure{f2}{} 
        \State start transaction
        \State f2's logic
        \State end transaction
    \EndProcedure

\end{algorithmic}

\end{algorithm}

This section presents \framework, a framework that allows smart contract
developers to write smart contracts with correct client transaction isolation semantics.
The goal of \framework is to provide developers with the primitives \textit{start
transaction} and \textit{end transaction}. We call each function surrounded by 
these primitives, a \textbf{transactional} function. \framework ensures that
calls to transactional functions are executed in isolation from any
concurrent function calls to the same function or any other function in the 
smart contract even in the presence of network asynchrony. 
Algorithm~\ref{algo:example-smart-contract} illustrates an example
smart contract written using \framework. This smart contract has two functions F1
and F2 and both functions are transactional functions.

The ACID execution of a client transaction requires \textit{atomic},
\textit{consistent}, \textit{isolated}, and \textit{durable}
execution of this client transaction. If the semantics of every smart 
contract function is correct, function calls should transfer the smart 
contract object from one consist state to another. Therefore, 
\textit{consistency} is the responsibility of the smart contract 
developer. \textit{Durability} of a function call is guaranteed
through the blockchain protocol. Function calls that complete execution
and are included in a mined block are durable assuming this block gets
enough confirmations~\cite{confirmation}. Since confirmed blocks are
replicated to most of the mining nodes, these blocks are durable even
in the presence of failures of many mining nodes. This leaves the
responsibility of ensuring atomicity and isolation of client transactions on \framework.

\textbf{Isolation:} Since smart contract developers have no way to 
detect which attribute values have been read by the client before a 
function call request is sent to miners, a smart contract developer has 
to insert checks at the beginning of every
smart contract function call (similar to optimistic concurrency 
control~\cite{kung1981optimistic})
to ensure that any data attribute value read by the client and is in the
read set of the function call matches its current value in the 
blockchain. The read set of a smart contract function is the set of attributes 
that a function reads during its execution. We assume that the outcome of each 
function is \textbf{invariant} to any attribute outside the read set of this function.
To ensure 
serializability~\cite{bernstein1987concurrency} of client transactions,
the client has to send her observed attribute values of the read set of 
the function along with the function call. The smart contract
has to ensure that the received attribute values are up-to-data and they
match the current values of all the attributes in the function read set
before executing the function call. Otherwise, the function call has to 
abort.  A function call and all its asynchronous callbacks
must be executed in isolation from concurrent function calls and 
callbacks.

\textbf{Atomicity:} The smart contract code has to
guarantee that a function call and all its asynchronous callbacks are 
atomic. This means that updates that result from a function call and all
its asynchronous callbacks should either all take place or none of them 
do. 

\framework automatically adds transaction isolation checks at the beginning of every transactional 
function to ensure an isolated execution of every call to any 
transactional function. \framework handles the atomicity of 
single domain transactional functions differently from cross-domain distributed 
transactional functions as follows.

\subsection{Single Domain Transactional Functions}

A Single Domain Transactional Function (SDTF for short) is a function 
that reads and updates one or more smart contract objects stored 
under a single administrative domain or a single blockchain. SDTFs 
do not access external services or
objects outside the domain of their blockchain. As a result,
SDTF calls do not trigger any 
asynchronous callbacks. Any transactional function that accesses
external services, blockchains, or trigger callbacks is classified as
cross-domain distributed transactional function.

Since all the objects accessed by SDTF calls are
stored in a miner's copy of the blockchain and since SDTFs do not 
trigger asynchronous callbacks, a SDTF call can atomically be executed 
in one shot. Therefore, the atomicity of a client transaction that
calls a SDTF is guaranteed by the smart contract execution model. To 
ensure a seriablizable execution of a SDTF, the function code has to 
only ensure the freshness of the read set of this function.
\framework scans every SDTF in a smart contract to determine the 
object's attributes in the read set of this SDTF. Then, \framework 
adds checks at the beginning of the SDTF to ensure that the attribute 
values observed by the client at the time when the transaction started
are equivalent to attribute values when the function call is received by
miners.

Recall the puzzle example in Algorithm~\ref{algo:puzzle-smart-contract}.
Both UpdateReward and SubmitSolution are single domain function calls.
To convert UpdateReward to a SDTF, \framework adds a requirement that 
every function call to the UpdateReward function must be accompanied by 
the client observed value of the attribute \textit{solved}, in the 
function read set, in its implicit parameter $msg.data$. Then,
\framework adds a requirement check $solved == msg.data.solved$. If the
$solvd$ attribute value in a client's UpdateReward function call is 
stale, the call must abort and the smart contract object state remains 
unchanged. However, if the $solved$ attribute is up-to-date and the 
function call is also accompanied by sufficient $gas$, the call can
be atomically executed and as a result, the reward value is updated.

For the SubmitSolution function call, \framework adds the requirement
checks $solved == msg.data.solved$ and $reward == msg.data.reward$. 
Recall the concurrency violation of the puzzle smart contract in 
Section~\ref{sec:examples}. When Bob sends
his solution to the SubmitSolution function, Bob would send the 
attribute values $solved = false$ and $reward = 2\:ethers$ in the 
$msg.data$ parameter of his function call. When Bob's request is 
received by a miner, there are two possible outcomes: 1) the function 
call gets executed only if the current reward value equals to 2 ethers 
and the puzzle is not solved and 2) the function call aborts if the 
reward value has been updated in between the time when Bob's transaction
started and the time when his function call is received by a miner. Both
outcomes do not violate the serializability 
guarantee.

\subsection{Cross-Domain Transactional Functions}

A Cross-Domain Distributed Transactional Function (CDTF for short) is a 
function that reads and updates one or more smart contract objects stored 
under multiple administrative domains or multiple blockchains. In addition, 
CDTFs can access external services or objects outside the domain of their 
blockchain. Also, CDTFs may trigger asynchronous callbacks. As a result,
updates made by a CDTF can span more than one block of the blockchain.
Recall the BlockKing smart contract in 
Algorithm~\ref{algo:blockking-smart-contract}. Each function call first
updates the warrior, warriorBlock, and warriorGold in some block and might
update the BlockKing in another block when the callback function is trigger.
Allowing a CDTF call to update the state of a smart contract object in 
several blockchain blocks is problematic. If the updates in the first 
block gets committed in a mined block, committed updates cannot be
rolled back even if updates in the following blocks fail due to an 
exception or that the call runs out of gas. We first explain isolation
and atomicity challenges of CDTFs. Afterwards, we explain how \framework
handles CDTFs.

\textbf{Isolation:} A CDTF has an entry point that comprises a function call
in an object in some blockchain. This function call might trigger other function 
calls of objects stored in different blockchains. Since a CDTF can span multiple 
blockchain objects, sending the read set of all the accessed objects at the entry 
point (the first function call) is not sufficient to guarantee transaction isolation 
as in SDFT. Since all subsequent function calls to other objects are trigger over 
an asynchronous network, the state of these subsequent objects might change in
the time between the entry point and the point when the subsequent call is received 
by the miners of the blockchain where these objects are stored. Even if the read set
is carried on with every subsequent call, a stale attribute in the read set might result
in aborting a subsequent call. However, the first call might have been committed leading
to a violation to atomicity. 

\textbf{Atomicity:} Guaranteeing the atomicity of CDTF calls is significantly more complicated
than SDFTs. First, atomicity could be violated if one of the subsequent calls to functions
in other blockchains runs out of $gas$. Second, if an external service (e.g., the random number
generator in the BlockKing example) crashes for a long time or if the message from 
this external services that triggers the callback function is lost, atomicity can be violated
resulting in an inconsistent state (some updates occur in one block but the callback is never
triggered to complete the execution of the function call).

Due to space limitation, we only present a high level solution that guarantees both
the isolation and atomicity of CDTFs. The atomic and isolated execution of a CDTF that
spans multiple blockchains can be mapped to the problem of atomic cross-chain transaction
processing. Atomic cross-chain commitment protocols have been introduced 
in~\cite{atomicNolan,herlihy2019cross,herlihy2018atomic,zakhary2019atomic}. First, the solution 
requires to lock all the object attributes in both the read set and the write set
of {\em all} the functions in a CDTF before 
calling the entry point. This locking guarantees the isolation of a CDTF from all concurrent 
function calls to any of the functions that can update either the read set or
the write set of a CDTF. However,
as shown in~\cite{zakhary2019atomic}, using timelocks as proposed 
in~\cite{atomicNolan,herlihy2018atomic} can lead to atomicity violations. The 
$AC^3WN$~\cite{zakhary2019atomic} and the $CBC$~\cite{herlihy2019cross} protocols that use
an additional blockchain as a lock manger are possible solutions to manage the locking
of object attributes across blockchains. After all the object attributes are locked, a caller can send a function
call request to the entry point accompanied 
by evidence that all the object attributes in both the read set and the write
set of this function call
and all subsequent function calls are locked. Object attributes are unlocked only when the function
call that accesses them and its corresponding callbacks, if any, terminate. 
Recall the BlockKing concurrency anomaly in Section~\ref{sec:examples}. Alice's call
locks the accessed attributes before calling the {\tt Enter} function. This prevents
other callers, Bob and Carol, from issuing concurrent function calls to the {\tt Enter} function.
Second, economic 
incentives should be used to enforce callers to accompany function calls with enough $gas$. 
At the entry point, a caller locks some money in the contract that gets refunded to the caller 
only if all her function calls terminate. If any function call runs out of gas, the caller 
loses her locked money to the contract owner who can complete the call and gets the locked 
objects unlocked. Finally, redo logs can be used to overcome the atomicity violations in the 
presence of external service crashes. In the BlockKing example, the smart contract object should
have an "after-image" attribute corresponding to every attribute in the 
object. The {\tt Enter} function
should update the after-images of warrior, warriorBlock, and warriorGold attributes. When the {\tt 
callback} is triggered, only then, the after-image attributes can be copied to the actual attributes of
the object. This guarantees that even if the external service crashes or the callback trigger  is lost, the object is in consistent state.
\section{Conclusion}\label{sec:conclusion}
In this paper, we presented \framework, a framework that allows developers to write
smart contracts with correct transactional semantics. We showed that \framework can help
developers solve isolation anomalies of both single domain and cross-domain distributed transactional functions.

\balance
\bibliographystyle{abbrv}
\bibliography{main}

\end{document}